\documentclass{Interspeech}
\interspeechcameraready

\title{Pairwise Evaluation of Accent Similarity in Speech Synthesis}

\author[affiliation={1}]{Jinzuomu}{Zhong}
\author[affiliation={2}]{Suyuan}{Liu}
\author[affiliation={1}]{Dan}{Wells}
\author[affiliation={1}]{Korin}{Richmond}

\affiliation{Centre for Speech Technology Research}{University of Edinburgh}{United Kingdom}
\affiliation{Department of Linguistics}{University of British Columbia}{Canada}
\email{\{jinzuomu.zhong, dan.wells, korin.richmond\}@ed.ac.uk, suyuan.liu@ubc.ca}

\newcommand{\blue}[1]{\textcolor{blue}{#1}}

\usepackage{comment}
\usepackage{multirow}
\begin{document}

\maketitle

\begin{abstract}
\vspace{-0.5em}
    Despite growing interest in generating high-fidelity accents, evaluating accent similarity in speech synthesis has been underexplored. We aim to enhance both subjective and objective evaluation methods for accent similarity. Subjectively, we refine the XAB listening test by adding components that achieve higher statistical significance with fewer listeners and lower costs. Our method involves providing listeners with transcriptions, having them highlight perceived accent differences, and implementing meticulous screening for reliability. Objectively, we utilise pronunciation-related metrics, based on distances between vowel formants and phonetic posteriorgrams, to evaluate accent generation. Comparative experiments reveal that these metrics, alongside accent similarity, speaker similarity, and Mel Cepstral Distortion, can be used. Moreover, our findings underscore significant limitations of common metrics like Word Error Rate in assessing underrepresented accents.
\end{abstract}

\begin{table*}[t]
\centering
\setlength{\tabcolsep}{1.4pt}
\captionsetup{justification=centering}
\caption{Evaluation methods used in various (accent) generation systems/papers. \blue{Blue} indicates objective evaluation methods.}
% : No reference speech is provided in these subjective evaluation tests. : These \blue{CER} are calculated on non-English languages, such as Mandarin. : Listeners are asked to classify different accent intensity levels rather than different accents in these AID tests.
\label{tab:lit_rev}
\vspace{-1em}
% \tiny
\scriptsize
\renewcommand{\arraystretch}{0.82}
\begin{tabular}{ll|lllllll}
\toprule
\multicolumn{1}{c}{Category} & \multicolumn{1}{c|}{System/Paper} & \multicolumn{1}{c}{Accent Similarity} & \multicolumn{1}{c}{Speaker Similarity} & \multicolumn{1}{c}{Intelligibility} & \multicolumn{1}{c}{Naturalness} & \multicolumn{1}{c}{Audio Quality} & \multicolumn{1}{c}{F0} & \multicolumn{1}{c}{Duration} \\ \midrule
\multirow{7}{*}{\begin{tabular}[c]{@{}l@{}}Zero-Shot\\ TTS\end{tabular}} 
 % & YourTTS, 2022 \cite{casanova2022yourtts} & & XMOS, \blue{CosSim} & & MOS & & & \\
 & XTTS, 2024 \cite{casanova2024xtts} & & XMOS, \blue{CosSim} & \blue{CER} & CMOS, \blue{UTMOS} & & & \\
 & VALL-E, 2025 \cite{chen2025neural} & & \blue{CosSim} & \blue{WER} & CMOS & & & \\
 % & Seed-TTS & & \blue{CosSim} & WER, CER & CMOS & & & \\
 & MaskGCT, 2025 \cite{wang2025maskgct} & XMOS, \blue{CosSim} & \blue{CosSim} & \blue{WER, CER} & MOS, CMOS & \blue{MCD, FSD} & & \\
 & Zhang et al., 2023 \cite{zhang2023towards} & XMOS(4), \blue{ AID-Acc} & XMOS(4) & & MOS & & \blue{RMSE, PCC, DTW} & \\
 & AccentBox, 2025 \cite{zhong2025accentbox} & XAB, \blue{CosSim} & XAB, \blue{CosSim} & & AB & & & \\ \midrule
\multirow{10}{*}{\begin{tabular}[c]{@{}l@{}}Accented\\ TTS\end{tabular}}
 & Zhou et al., 2024 \cite{zhou2024multi} & XAB, XMOS, \blue{CosSim} & XAB, XMOS, \blue{CosSim} & & AB, MOS & \blue{MCD} & \blue{RMSE, PCC} & \blue{FD} \\
 & Zhou et al., 2024 \cite{zhou2024accented} & XAB, XBWS, \blue{PPG-KL} & & & AB, MOS & & \blue{RMSE, PCC} & \blue{FD, RMSE} \\
 % & Ai-TTS, 2023 \cite{liu2023explicit} &  AID-Acc & & & MOS & & \blue{Moments, DTW} & \\
 & CTA-TTS, 2024 \cite{liu2024controllable} & XMOS,  AID-Acc & & & & MOS, BWS, \blue{MCD} & \blue{Moments, DTW} & \blue{MAE} \\
 % & TTS-MLVAE, 2022 \cite{melechovsky2022learning} & XBWS & XMOS, \blue{EER} & \blue{WER} & MOS & \blue{MCD} & & \\
 & DART, 2024 \cite{melechovsky2024dart} & XBWS & XBWS, \blue{CosSim} & \blue{WER} & MOS & \blue{MCD} & \blue{FFE} & \\
 & RAD-MMM, 2023 \cite{badlani2023rad} & XCMOS, \blue{CosSim} & XCMOS & \blue{CER} & & & & \\
 % & VANI, 2023 \cite{rohan2023vani} & & XMOS, \blue{CosSim} & \blue{CER} & MOS & & & \\
 & AccentSpeech, 2022 \cite{zhang2022accentspeech} & XAB & XAB, \blue{CosSim} & & AB & & & \blue{MAE}  \\
 & Deja et al., 2023 \cite{deja2023diffusion} & MUSHRA,  AID-Acc & XMUSHRA & & MUSHRA & & & \\
 & Yang et al., 2023 \cite{yang2023parameter} & MOS & & \blue{CER} & MOS & \blue{MCD} & & \\ \midrule
\multirow{10}{*}{\begin{tabular}[c]{@{}l@{}}Accent\\ Conversion\end{tabular}}
 % & Zhao et al., 2018 \cite{zhao2018accent} & AB & XMOS(15) & & & MOS & & \\
 % & Zhao et al., 2019 \cite{zhao2019foreign} & MOS(9) & XAB & & MOS & MOS & & \\
 & Accentron, 2022 \cite{ding2022accentron} & MOS(9) & XMOS(15) & & & MOS & & \\
 & Quamer et al., 2022 \cite{quamer2022zero} & MOS(9) & XAB & & MOS & & & \\
 & Zhou et al., 2023 \cite{zhou2023tts} & MOS(9) & XMOS(4) & \blue{WER, PER} & & MUSHRA & & \\
 & Bai et al., 2024 \cite{bai2024diffusion} & XBWS & XBWS & & MUSHRA & & & \\
 & Chen et al., 2024 \cite{chen2024transfer} & MOS & XAB & \blue{WER} & MOS & & & \\
 & Jia et al., 2023 \cite{jia2023zero} & XMOS & XMOS & & MOS & & & \\
 & Jin et al., 2023 \cite{jin2023voice} & XAB & XMOS & & & MOS & & \\
 & Jia et al., 2024 \cite{jia2024convert} & AID-Acc, \blue{AID-Acc} & \blue{CosSim} & & MOS, \blue{NISQA} & & & \\
 & MacST, 2025 \cite{inoue2025macst} & MUSHRA, \blue{CosSim} &  \blue{CosSim} & \blue{WER} & MUSHRA & & & \\ \bottomrule
\end{tabular}
\caption*{\scriptsize{X...: Reference speech is provided for similarity ratings. (\texttt{num}): A \texttt{num}-point rather than 5-point scale is used. \blue{AID-Acc}: accent identification accuracy. \blue{CosSim}: cosine similarity of accent/speaker embeddings. \blue{PPG-KL}: KL divergence of phonetic posteriorgrams. \blue{WER/CER/PER}: word/character/phone error rate. \blue{UTMOS / NISQA}: predicted MOS by \cite{saeki2022utmos} / \cite{mittag2020deep}. \blue{MCD}: mel cepstral distortion. \blue{FSD}: Fréchet speech distance. \blue{RMSE}: root mean square error. \blue{PCC}: Pearson correlation coefficient. \blue{DTW}: DTW alignment cost. \blue{Moments}: see \cite{ren2021fastspeech}. \blue{FFE}: F0 frame error. \blue{FD}: frame difference. \blue{MAE}: mean absolute error.}}
\vspace{-4em}
\end{table*}

\vspace{-1em}
\section{Introduction}
\vspace{-0.5em}

Motivated by the social and moral imperative for more inclusive speech technology, the community has witnessed a growing interest in systems capable of generating high-fidelity \textit{accent}, across various speech generation tasks. In Zero-Shot Text-to-Speech (ZS-TTS), accent hallucination/mismatch, where the generated speech deviates from the reference speech in accent, is reported in \cite{chen2025neural, wang2025maskgct, zhong2025accentbox} and addressed in \cite{wang2025maskgct, zhang2023towards, zhong2025accentbox}. In Accented TTS, numerous attempts have been made to generate high-fidelity accent based on pre-defined accent variety labels or intensity levels \cite{zhou2024multi}--\cite{yang2023parameter}. In Accent Conversion (AC), numerous attempts have been made to map speech from foreign (L2) to native (L1) accent, preserving content and speaker information while removing the foreign accent in the source speech \cite{ding2022accentron}--\cite{inoue2025macst}. However, how to evaluate accent similarity in speech is an under-researched topic that lacks consensus. Tab.\ \ref{tab:lit_rev} shows a non-exhaustive list of various (accent) generation systems/papers, along with the subjective and objective methods they employ for evaluation.

For \textit{subjective} evaluation, there are mainly two categories: \textit{reference-free} and \textit{reference-based} listening tests. \textit{Reference-free} methods ask listeners to rate speech utterances without any reference speech. Listeners are tasked to rate the degree of L2 accent, such as the accentedness test \cite{munro1995foreign} commonly used in AC \cite{ding2022accentron, quamer2022zero, zhou2023tts} or to rate the resemblance to a pre-defined accent label such \cite{deja2023diffusion, yang2023parameter, chen2024transfer}. These methods wrongly assume accents to be on a one-dimensional scale of intensity, or categorical (only comprising standard accent varieties) rather than varying on an individual basis \cite{markl2023everyone}. \textit{Reference-based} methods borrow the AB/MOS/CMOS listening tests from standard evaluation of naturalness or audio quality, and ask listeners to rate the accent similarity between reference speech X and target speech utterances, denoted as XAB/XMOS/XCMOS. However, no prior work has studied the validity of these tests specifically on evaluating accent similarity, with most relevant work focusing on test designs in evaluating naturalness \cite{camp2023mos, wells2024experimental}.

\vspace{-0.25em}
In the absence of reliable and cost-effective subjective evaluation methods, many systems/papers turn to \textit{objective} evaluation, mostly using either classification results or cosine similarity of embeddings extracted from Accent Identification (AID) models. However, how well these proxy metrics represent accent similarity remains unstudied, with relevant work focusing on the correlation between objective and subjective evaluation of speaker similarity \cite{liu2024a}. Meanwhile, these systems/papers report a broad range of objective metrics to proxy intelligibility, naturalness, and so on (see Tab.\ \ref{tab:lit_rev}). However, some of these methods may be unfair for evaluating speech of underrepresented accents, namely the accent bias in WER \cite{sanabria2023the}.

\vspace{-0.25em}
Given this research gap, we explore the question: \textit{How should accent similarity be evaluated both subjectively and objectively?} For subjective evaluation, due to the reliability of AB tests and the limitations found with MOS tests \cite{camp2023mos, wells2024experimental, le2024limits}, we build upon the XAB listening test, with three additional components that achieve higher statistical significance with fewer listeners and lower costs.
Our method involves providing listeners with transcriptions, having them highlight perceived accent differences, and implementing meticulous screening for reliability.
For objective evaluation, inspired by \cite{zhou2024accented, huckvale2004accdist, churchwell2024high}, we propose to use pronunciation-related metrics based on distances between vowel formants and phonetic posteriorgrams (PPGs), both of which capture phonetic identities, to evaluate accent generation. We also comparatively experiment with a broad range of objective metrics and their correlation to the ranking of several systems (with hypothesised different qualities of accent generation). Our findings show these pronunciation-related metrics, alongside cosine similarity of Accent Identification (AID) or Speaker Verification (SV) embeddings, and Mel Cepstral Distortion (MCD), can be used for evaluating speech of underrepresented accents. Moreover, our findings underscore significant limitations of metrics like WER in assessing these underrepresented accents, calling for more research into fair and inclusive speech evaluation. To summarise, our contributions are:
\vspace{-0.5em}
\begin{itemize}
    \item We propose several refinement strategies to make subjective evaluation of accent similarity more reliable.
    \vspace{-0.25em}
    \item We propose to utilise pronunciation-related metrics, based on distances between vowel formants and PPGs, to objectively evaluate accent similarity in speech.
    \vspace{-0.25em}
    \item Comparative experiments show that our refined XAB subjective evaluation, proposed pronunciation metrics, and several other objective metrics, can be used for evaluating underrepresented accents.
\end{itemize}

\section{Subjective Evaluation}
\label{sec:subj}
\vspace{-0.5em}

\subsection{XAB accent similarity listening test}
\label{ssec:xab_baseline}
\vspace{-0.5em}

In our listening tests, listeners are instructed as follows:
\textit{``Listen carefully to all speech recordings below in full. Then pick the candidate speech recording that is \textit{more similar} in terms of \textit{accent} to the reference speech recording. Please disregard the mismatch in voice, gender, and audio quality.''}
We ensure that the reference and candidate speech share identical content but are delivered by two different speakers of different genders. Having identical content allows listeners to focus on detailed pairwise comparisons of pronunciation and prosody, essential for assessing accent \cite{huckvale2004accdist}. By rating the similarity of utterances from speakers of different genders, listeners are prevented from confusing speaker similarity with accent similarity.

\vspace{-1em}
\subsection{Highlight of perceived accent difference}
\label{ssec:xab_highlight}
\vspace{-0.5em}

Inspired by Rapid Prosody Transcription \cite{gutierrez2021location}, we propose to additionally provide the text transcript of test utterances to listeners, and ask them to highlight the perceived accent differences. This should further guide listeners to nuanced differences related to accents. The detailed instructions are as follows:
\textit{``i) Please carefully highlight only the specific parts of the sentence that helped you decide how similar one recording's accent is to another.
ii) Avoid selecting the entire sentence. Instead, try to be as precise as possible. For example, if the `tt' sound in `bottle' influenced your decision, highlight just `tt'.
iii) Click and drag to select or deselect parts of the text. If you make a mistake, you can click again to undo your selection.
iv) Use the `Clear All Highlights' button below to remove all highlights if you wish to start over."}

\vspace{-1em}
\subsection{Screening listeners}
\label{ssec:xab_screen}
\vspace{-0.5em}

To screen out invalid submissions, we devise two components. First, we include several attention-check questions with speech samples, one (A) from the same accent as in the reference (X), the other (B) from a very different accent, expecting the listeners to select A. Second, we include an open-ended accent identification question at the end, requesting listeners to identify the accent in the reference speech as specifically as possible. We expect listeners to be able to identify at least the country if not the city where the accent is commonly spoken.

% \begin{table}[h!]
% \captionsetup{justification=centering}
% \caption{Objective metrics used in experiments.}
% \label{tab:obj}
% \vspace{-0.5em}
% \begin{tabular}{ll}
% \toprule
% \multicolumn{1}{c}{Category} & \multicolumn{1}{c}{Metrics} \\ \midrule
% Pronunciation & Vow. For., PPG CosSim, PPG JS \\
% Accent Similarity & ComAcc CosSim, GenAID CosSim \\
% Speaker Similarity & WavLM CosSim \\
% Intelligibility & Whisper WER, Whisper CER \\
% Naturalness & NTMOS \\
% Audio Quality & MCD \\
% F0 & F0 RMSE, F0 Per. RMSE, F0 PCC \\ \bottomrule
% \end{tabular}
% \end{table}

\begin{table*}[t]
\centering
\captionsetup{justification=centering}
\caption{Results of different objective evaluation metrics and their correlation with hypothesised ranking (Hyp.\ Rank), $\dagger$: p $>$ 0.05, i.e.\ not statistically significant. VF: vowel formants. JS: Jensen-Shannon distance. SRCC: Spearman rank correlation coefficient.}
\label{tab:obj_metrics}
\vspace{-1em}
\scriptsize
\setlength{\tabcolsep}{2pt}
\begin{tabular}{lcccccccccccccc}
\toprule
% \multicolumn{1}{c}{\multirow{4}{*}{System}} & \multicolumn{1}{c}{\multirow{4}{*}{\begin{tabular}[c]{@{}c@{}}Hyp.\\ Rank\end{tabular}}} & \multicolumn{3}{c}{Pronunciation} & \multicolumn{2}{c}{\begin{tabular}[c]{@{}c@{}}Accent\\ Similarity\end{tabular}} & \multicolumn{1}{c}{\begin{tabular}[c]{@{}c@{}}Speaker\\ Similarity\end{tabular}} & \multicolumn{2}{c}{Intelligibility} & \multicolumn{1}{c}{\begin{tabular}[c]{@{}c@{}}Natural\\ -ness\end{tabular}} & \multicolumn{1}{c}{\begin{tabular}[c]{@{}c@{}}Audio\\ Quality\end{tabular}} & \multicolumn{3}{c}{F0} \\ \cmidrule{3-15}
\multicolumn{1}{c}{\multirow{4}{*}{System}} & \multicolumn{1}{c}{\multirow{4}{*}{\begin{tabular}[c]{@{}c@{}}Hyp.\\ Rank\end{tabular}}} & \multicolumn{3}{c}{Pronunciation} & \multicolumn{2}{c}{Accent Similarity} & \multicolumn{1}{c}{Speaker Similarity} & \multicolumn{2}{c}{Intelligibility} & \multicolumn{1}{c}{Naturalness} & \multicolumn{1}{c}{Audio Quality} & \multicolumn{3}{c}{F0} \\ \cmidrule{3-15}
\multicolumn{1}{c}{} & \multicolumn{1}{c}{} & \multicolumn{1}{c}{\begin{tabular}[c]{@{}c@{}}VF\\ RMSE$\downarrow$\end{tabular}} & \multicolumn{1}{c}{\begin{tabular}[c]{@{}c@{}}PPG\\ CosSim$\uparrow$\end{tabular}} & \multicolumn{1}{c}{\begin{tabular}[c]{@{}c@{}}PPG\\ JS$\downarrow$\end{tabular}} & \multicolumn{1}{c}{\begin{tabular}[c]{@{}c@{}}GenAID\\ CosSim$\uparrow$\end{tabular}} & \multicolumn{1}{c}{\begin{tabular}[c]{@{}c@{}}ComAcc\\ CosSim$\uparrow$\end{tabular}} & \multicolumn{1}{c}{\begin{tabular}[c]{@{}c@{}}WavLM\\ CosSim$\uparrow$\end{tabular}} & \multicolumn{1}{c}{\begin{tabular}[c]{@{}c@{}}Whisper\\ WER$\downarrow$\end{tabular}} & \multicolumn{1}{c}{\begin{tabular}[c]{@{}c@{}}Whisper\\ CER$\downarrow$\end{tabular}} & \multicolumn{1}{c}{UTMOS$\uparrow$} & \multicolumn{1}{c}{MCD$\downarrow$} & \multicolumn{1}{c}{\begin{tabular}[c]{@{}c@{}}F0\\ RMSE$\downarrow$\end{tabular}} & \multicolumn{1}{c}{\begin{tabular}[c]{@{}c@{}}F0 Per.\\ RMSE$\downarrow$\end{tabular}} & \multicolumn{1}{c}{\begin{tabular}[c]{@{}c@{}}F0\\ PCC$\uparrow$\end{tabular}} \\ \midrule
\texttt{copysyn} & 1 & 62.12 & 0.9744 & 0.0684 & 0.9927 & 0.9558 & 0.9443 & 1.737 & 0.724 & 3.732 & 2.800 & 299.1 & 0.01076 & 0.8138 \\
\texttt{xtts} & 2 & 177.98 & 0.8793 & 0.1936 & 0.9447 & 0.7792 & 0.9248 & 1.262 & 0.499 & 3.912 & 5.571 & 482.7 & 0.01296 & 0.6164 \\
\texttt{corrupt30k} & 3 & 217.96 & 0.8717 & 0.1981 & 0.8399 & 0.7215 & 0.8778 & 0.364 & 0.151 & 4.108 & 6.067 & 431.2 & 0.01135 & 0.6427 \\
\texttt{corrupt60k} & 4 & 212.47 & 0.8706 & 0.1992 & 0.8403 & 0.7108 & 0.8758 & 1.209 & 0.839 & 4.115 & 6.093 & 456.9 & 0.01045 & 0.6027 \\
\texttt{corrupt90k} & 5 & 225.88 & 0.8594 & 0.2072 & 0.8302 & 0.7089 & 0.8726 & 1.896 & 1.578 & 4.121 & 6.232 & 441.6 & 0.01132 & 0.6316 \\
\texttt{corrupt120k} & 6 & 224.04 & 0.8388 & 0.2241 & 0.8394 & 0.7104 & 0.8722 & 3.909 & 3.845 & 4.085 & 6.658 & 425.6 & 0.01077 & 0.6695 \\
\texttt{corrupt150k} & 7 & 230.67 & 0.8404 & 0.2225 & 0.8369 & 0.7095 & 0.8676 & 2.331 & 2.091 & 4.104 & 6.576 & 441.6 & 0.01008 & 0.6272 \\ \midrule
% \multicolumn{2}{c}{PCC w/ Hyp.\ Rank} & 0.7749 & 0.8317 & 0.7621 & 0.8078 & 0.7505 & 0.8575 & 0.5963 & 0.7442 & 0.7656 & 0.7902 & 0.4242 & 0.5300 & 0.4983 \\[-0.1em]
% \multicolumn{2}{c}{p-value} & 0.0407 & 0.0203 & 0.0464 & 0.0280 & \underline{0.0520} & 0.0136 & \underline{0.1576} & \underline{0.0551} & 0.0448 & 0.0345 & \underline{0.3428} & \underline{0.2211} & \underline{0.2551} \\[0.2em]
\multicolumn{2}{c}{SRCC w/ Hyp.\ Rank} & 0.9286 & 0.9643 & 0.9643 & 0.8571 & 0.8929 & 1.0000 & 0.6429 & 0.8214 & 0.4643 & 0.9643 & 0.1071 & 0.4643 & 0.1786 \\[-0.1em]
\multicolumn{2}{c}{p-value} & 0.0025 & 0.0005 & 0.0005 & 0.0137 & 0.0068 & 0.0000 & 0.1194$^\dagger$ & 0.0234 & 0.2939$^\dagger$ & 0.0005 & 0.8192$^\dagger$ & 0.2939$^\dagger$ & 0.7017$^\dagger$ \\ \bottomrule
\end{tabular}
\vspace{-2em}
\end{table*}

\section{Objective Evaluation}
\label{sec:obj}
\vspace{-0.5em}

\subsection{Commonly used objective metrics}
\vspace{-0.5em}

We first extract several representative and commonly used objective metrics in evaluating TTS/VC systems. To evaluate accent/speaker similarity, we calculate the cosine similarity of accent/speaker embeddings (\texttt{CosSim}) using two AID models: 1) CommonAccent (\texttt{ComAcc}) \cite{zuluagagomez2023commonaccent}, 2) \texttt{GenAID} \cite{zhong2025accentbox}, and one SV model: 3) \texttt{WavLM-base-plus-sv} \cite{chen2022WavLMLargeScaleSelfSupervised}. To evaluate intelligibility, we calculate the WER/CER using Whisper\footnote{\url{https://huggingface.co/openai/whisper-large-v3}} with reference to ground truth transcription. To evaluate naturalness, we use \texttt{UTMOS} \cite{saeki2022utmos}. To evaluate audio quality and F0 related metrics, we use the Amphion \cite{zhang2024amphion} evaluation pipeline to extract MCD, F0 RMSE, F0 Periodicity RMSE, and F0 Pearson Correlation Coefficient (PCC).

\vspace{-1em}
\subsection{Vowel Formant Distance}
\vspace{-0.5em}

% vocal tract length normalisation by \cite{johnson2020deltaf}

Vowel formants are extracted as a reasonably interpretable evaluation metric. The first and second formants (F1\&F2) roughly correspond to vowel height and frontness, respectively \cite{wells1982accents}. We generate phone-level alignments using Montreal Forced Aligner \cite{mcauliffe2017MontrealForcedAligner} with pretrained \texttt{english\_us\_arpa} acoustic model/dictionary, and extract formants using Fast Track \cite{barreda2021fast} estimation at vowel midpoints. Finally, we calculate pairwise vowel formants RMSE (\texttt{VF RMSE}) using the extracted F1\&F2.

\vspace{-1em}
\subsection{Pronunciation distance based on DTW-aligned PPGs}
\vspace{-0.5em}

Mainly inspired by \cite{churchwell2024high} which proposes using normalised Jensen-Shannon distance between PPGs as pronunciation distance, we propose to use PPGs to measure accent-related pronunciation distance. We first extract the PPGs following \cite{churchwell2024high}, then additionally align the PPGs using Dynamic Time Warpping (DTW) alignment, to handle mismatch in audio length, serving both TTS and VC tasks. Different from \cite{churchwell2024high}, we drop the normalisation, since the normalisation in their work is biased by what General American phones are similar, based on the accent-imbalance data/model. We experiment with two alignment cost functions: cosine similarity distance and Jensen-Shannon distance, treating PPGs as features and probability distributions respectively. We calculate the mean cost over all alignment steps as the final pronunciation distance of the pairwise comparison.

\vspace{-1em}
\section{Experiments}
\vspace{-0.5em}

\subsection{Subjective evaluation}
\vspace{-0.5em}

\noindent \textbf{Test stimuli:} We use utterances 001--023 from VCTK\footnote{\url{https://datashare.ed.ac.uk/handle/10283/
3443}}, i.e.\ the rainbow passage and elicitation paragraph read by all speakers, as our test stimuli. Utterance 024 is arbitrarily chosen as speech prompt, for inferencing ZS-TTS systems.

\noindent \textbf{Systems:} In all XAB listening tests, the ground truth audio is used as the reference (X). The \texttt{copysyn} system (A) involves passing the mel-spectrogram from the ground truth to the pretrained HiFi-GAN model \cite{kong2020hifigan}. The \texttt{xtts} \cite{casanova2024xtts} system (B) generates audio from text and a speech prompt. As \texttt{xtts} lacks explicit accent control and produces noticeable accent errors, while \texttt{copysyn} merely reconstructs the audio with some vocoder artefacts, we hypothesise that listeners should prefer \texttt{copysyn} over \texttt{xtts} with reference to ground truth. % Due to budget constraints, we limit our XAB tests to these two systems but believe the findings would extend to other system comparisons.

\noindent \textbf{Test accent \& speaker:} We focus on the Edinburgh accent, typically underrepresented in English TTS/VC systems. From VCTK's five self-reported Edinburgh accent speakers (1 female, 4 male), we select the female speaker p262 as reference (X). A pilot study identified the male speaker p252 as most closely resembling the accent of the reference speaker, so we select p252 to be the test speaker in evaluated systems (A \& B)

\noindent \textbf{Listening tests:} We conduct three primary listening tests: baseline \texttt{XAB} as detailed in Sec.\ \ref{ssec:xab_baseline}, \texttt{XAB+trans} (with text transcription shown), and \texttt{XAB+trans+highlight} (highlighting perceived accent differences, as per Sec.\ \ref{ssec:xab_highlight}). Each of these initial tests involves 15 listeners. We further screen out invalid submissions using criteria from Sec.\ \ref{ssec:xab_screen} until we reach 15 \emph{valid} submissions, leading to additional sets of test results: \texttt{XAB+screen}, \texttt{XAB+trans+screen}, and \texttt{XAB+trans+highlight+screen}. Listeners in all listening tests are recruited on \texttt{Prolific} (paid £12/hour) with self-reported normal hearing, and birthplace \& current residence in the test accent region (i.e.\ Scotland). To avoid learning effects, there is no overlap in listeners between different listening tests.

\noindent \textbf{Statistical testing:} We employ one-sided t-tests with the null hypothesis that \texttt{copysyn} is not preferred over \texttt{xtts}. Due to the correlation between ratings across multiple utterances by a single listener, we calculate 95\% confidence intervals (CI) across multiple listeners/submissions rather than multiple utterances to avoid underestimates \cite{camp2023mos}. We also randomly sample a subset from all valid submissions, and calculate the expected statistical significance (p-value) of these subsets with respect to the number of valid submissions.

\vspace{-1em}
\subsection{Objective evaluation}
\vspace{-0.5em}

\noindent \textbf{Additional systems}: Objective evaluation allows for more systems to be evaluated, as it avoids costly listening tests. In addition to \texttt{copysyn} and \texttt{xtts}, we introduce a series of corrupted systems which are finetuned from original \texttt{xtts} on a single-speaker General American corpus, LJ~Speech\footnote{\url{https://keithito.com/LJ-Speech-Dataset/}}, with a batch size of 252 and learning rate of 5e-6. We hypothesise that longer fine-tuning leads to catastrophic forgetting, worsening non-General American accents. These corrupted systems are denoted as \texttt{corrupt30k}, \texttt{corrupt60k}, \texttt{corrupt90k}, \texttt{corrupt120k}, and \texttt{corrupt150k}, with finetuning conducted over steps ranging from 30k to 150k.

\noindent \textbf{Evaluation methods:} We use the same test stimuli as in subjective evaluation, with all speakers in Edinburgh accent. We calculate the mean of each objective metric listed in Section \ref{sec:obj}.

\vspace{-1em}
\section{Results \& Analysis}
\vspace{-0.5em}

\subsection{Subjective Evaluation}
\vspace{-0.5em}

\begin{figure}[h!]
\vspace{-1.5em}
\centering
\captionsetup{justification=centering}
\includegraphics[width=0.8\linewidth, height=0.65\linewidth, trim = 28 9 43 43, clip]{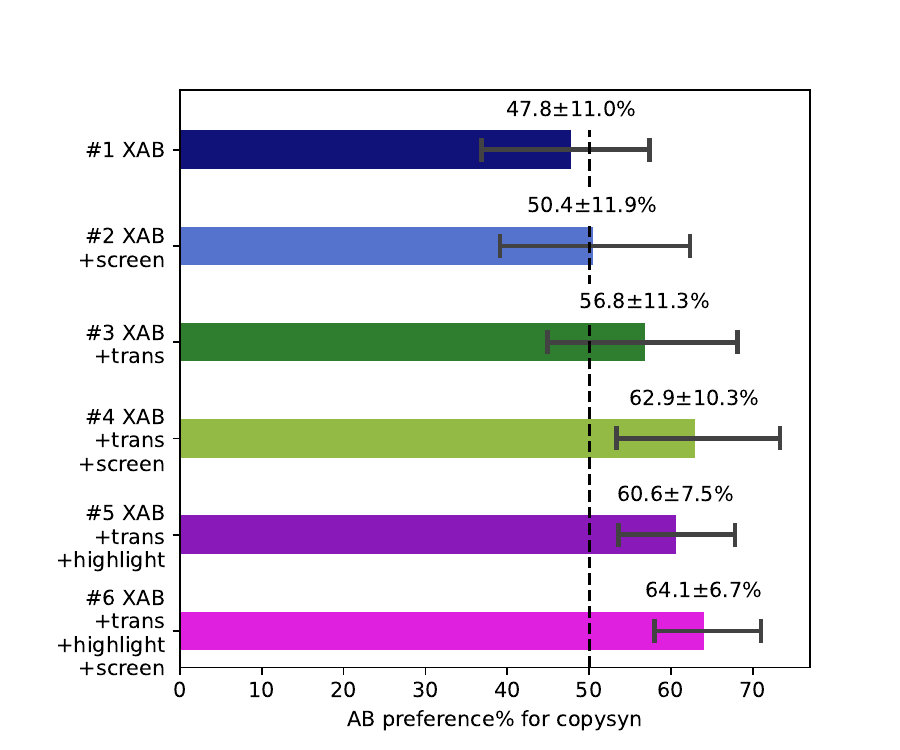}
\vspace{-1.2em}
\caption*{\scriptsize(a) AB preference\% (mean±95\%CI) of 15 valid submissions.}
\vspace{0.5em}
\includegraphics[width=0.8\linewidth, height=0.7\linewidth, trim = 12 19 42 51, clip]{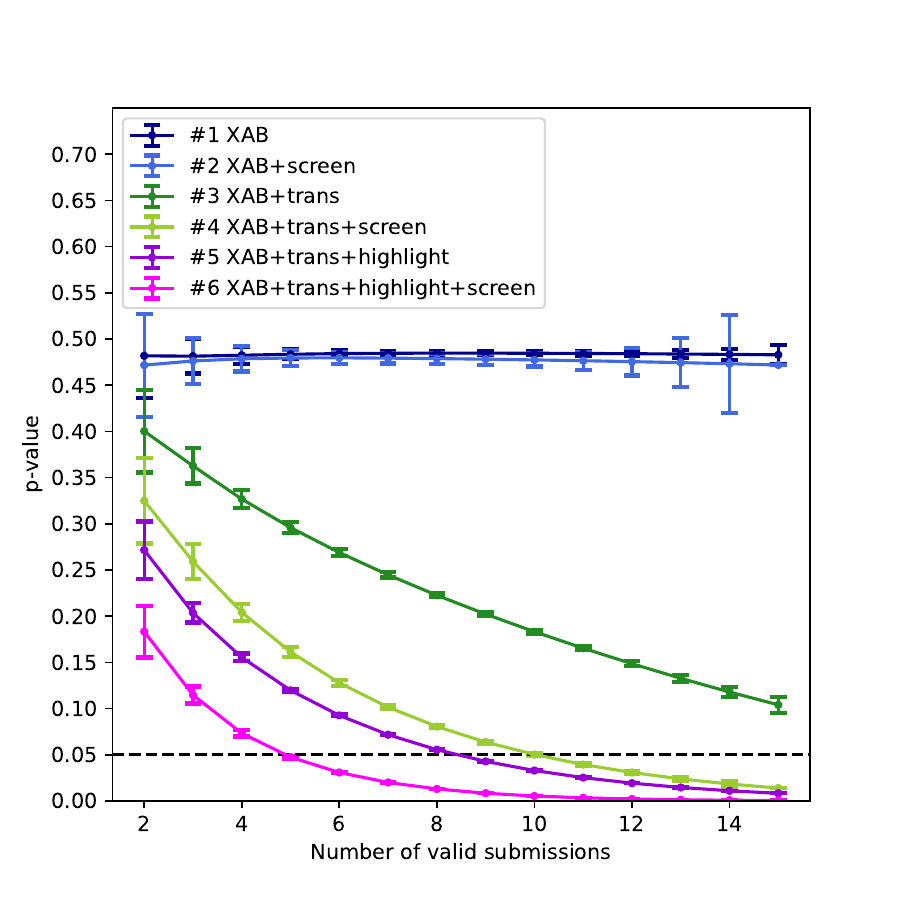}
\vspace{-1.2em}
\caption*{\scriptsize(b) p-value (mean±95\%CI) of random subsets of valid submissions.}
\vspace{-0.8em}
\caption{Comparison of different accent similarity XAB listening tests (\texttt{copysyn} vs \texttt{xtts}), with final results shown in (a) and statistical significance investigated in (b).}
\label{fig:subj_results}
\vspace{-1em}
\end{figure}

\noindent \textbf{AB Preference\% and statistical significance:}
The effects of \texttt{+trans}, \texttt{+high\-light}, and \texttt{+screen} are shown in Fig.~\ref{fig:subj_results}. Notably, test designs \texttt{\#1} and \texttt{\#2} do not reveal a preference for \texttt{copysyn}, evidenced by below/near 50\% preference in Fig.~\ref{fig:subj_results}(a), and constantly high p-values in Fig.~\ref{fig:subj_results}(b). This suggests that naive XAB listening tests, even with listeners screening (\texttt{+screen}), cannot meaningfully distinguish two systems in accent similarity. Providing listeners with transcription (\texttt{+trans}) in tests \texttt{\#3} and \texttt{\#4} leads to a preference for \texttt{copysyn}, with statistically significant preference achieved in test \texttt{\#4}, i.e.\ p-value dropping below 0.05. This suggests that listeners are more attentive to the accent-related differences between utterances when provided with reference transcriptions. Statistical significance can be achieved by either careful screening within 10 valid submissions or potentially by more than 15 valid submissions (see the p-values of system \texttt{\#3} that seem on track to drop below 0.05). Requesting listeners to complete an auxiliary task of highlighting accent-related differences (\texttt{+highlight}) in tests \texttt{\#5} and \texttt{\#6} leads to both higher preference for \texttt{copysyn} and higher statistical significance. The most effective listening test design \texttt{\#6} requires as few as 5 valid submissions to reach statistical significance, and reaches the highest 64.1±6.7\% preference among all listening tests.

\noindent \textbf{Duration, rejection rate, and cost analysis:}
On average, participants take 16.9±3.4 / 18.5±2.6 / 30.6±7.7 minutes to complete listening tests \texttt{\#2} / \texttt{\#4} / \texttt{\#6}, respectively. For the duration analysis only, we exclude data from 1 participant who has spent more than two standard deviations above the mean time. Screening listeners (\texttt{+screen}) minimally impacts duration, therefore duration statistics of \texttt{\#1} / \texttt{\#3} / \texttt{\#5} listening tests are not reported.
Providing listeners with transcription (\texttt{+trans}) does not increase the duration much; however, the auxiliary task of highlighting accent-related differences (\texttt{+highlight}) nearly doubles the time required to complete the test. This indicates that for accents with enough participants, the statistical significance gain of \texttt{+highlight} is offset by longer completion time, resulting in a similar budget cost; for accents with a limited pool of available participants, \texttt{+highlight} still remains an effective method.
Screening listeners (\texttt{+screen}) leads to a rejection rate of 16.7\% / 11.8\% / 25.0\% for listening tests \texttt{\#2} / \texttt{\#4} / \texttt{\#6}, respectively, incurring slightly higher costs. The rejection rate, probably, has less to do with test design, but more to do with the luck of a random sample from the Prolific participants. All ``rejected'' listeners pass the attention-check questions, but fail the AID question: they either cannot tell what the accent of the reference speech is (answering e.g.\ ``not sure'') or misidentifying it to be a non-Scottish accent (answering e.g.\ ``Southern England'').

\vspace{-1em}
\subsection{Objective Evaluation}
\vspace{-0.5em}

\begin{figure}[h!]
    \centering
    \includegraphics[width=0.75\linewidth]{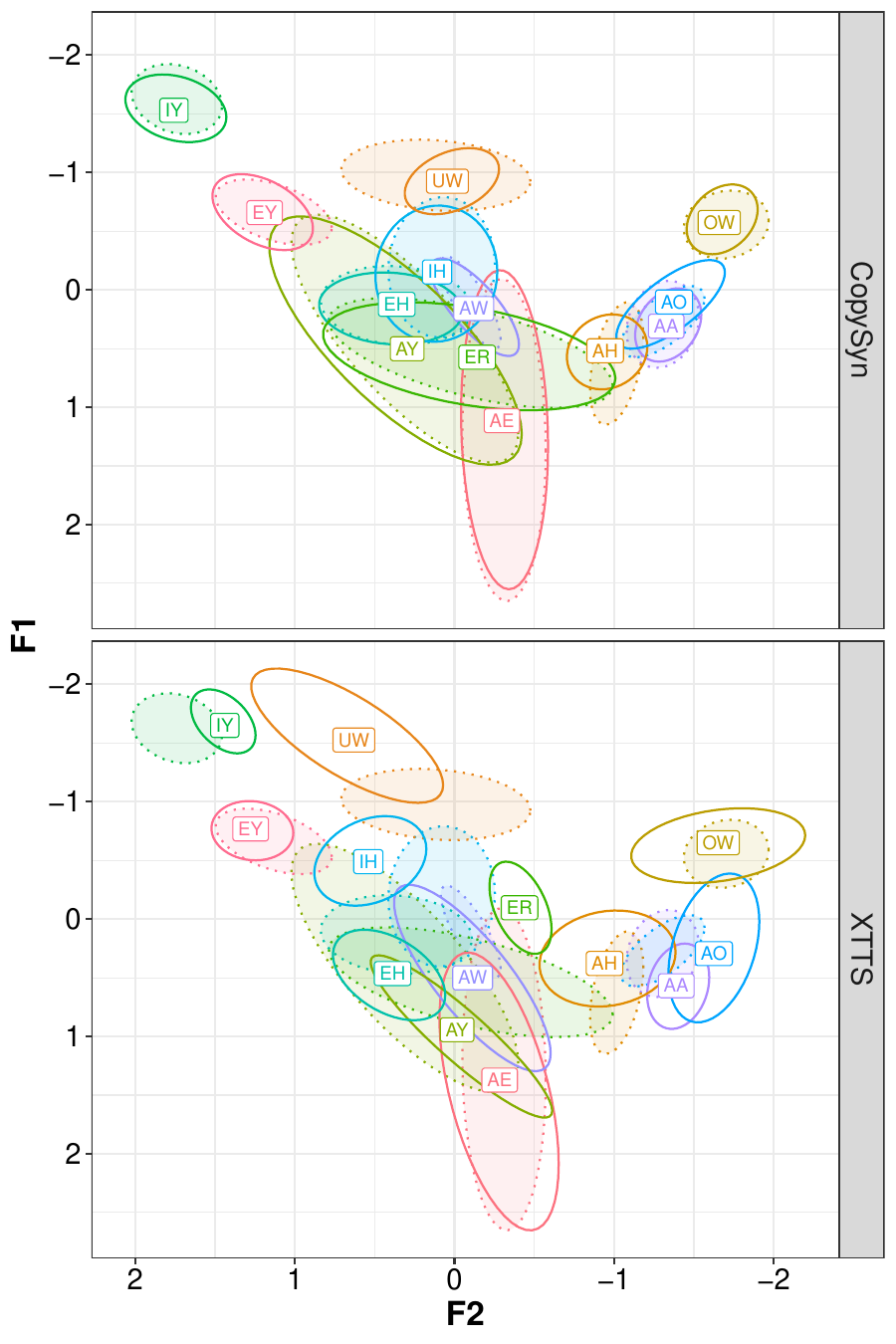}
    \vspace{-1.5em}
    \captionsetup{justification=centering}
    \caption{Disparity of formant distribution between ground truth (shaded ellipses, dotted lines) and \texttt{copysyn}/\texttt{xtts} (hollow ellipses, solid lines) for speaker p252. Vowel symbols are ARPABET. F1/F2 axes are normalised for each speaker.}
    \label{fig:obj_vowelspace}
    \vspace{-2em}
\end{figure}

\noindent The results of all objective metrics and their correlation with the hypothesised ranking of various synthesis systems are shown in Tab.~\ref{tab:obj_metrics}. The three proposed pronunciation metrics, alongside accent similarity, speaker similarity, and MCD metrics, exhibit strong correlations with the hypothesised ranking (all SRCCs $>$ 0.85 with statistical significance). However, the high correlation of speaker similarity may be influenced by factors beyond purely accent characteristics, given the catastrophic forgetting introduced by our corruption scheme, which impacts both accent and unseen speaker modelling. This interaction warrants further investigation. The unreliability of WER/CER (where \texttt{copysyn} is worse than \texttt{xtts}, and \texttt{xtts} worse than \texttt{corrupted30k}, despite p $<$ 0.05 for CER) likely stems from inherent accent bias in these models. Similarly, UTMOS is possibly inconsistent due to accent biases in the training data used for MOS prediction. Therefore, we recommend against using WER/CER/UTMOS metrics in evaluating accent generation. F0 metrics correlate poorly with the hypothesized ranking across different systems, likely because the corruption affects acoustic modelling aspects other than F0. Finally, we visualise the vowel space for p252, the test speaker, in Fig.\ \ref{fig:obj_vowelspace}, with near-perfect overlapping of vowel space/distribution between ground truth and \texttt{copysyn} and clear mismatching between ground truth and \texttt{xtts}. Unlike the ground truth and \texttt{copysyn}, the \texttt{xtts} vowel space appears to avoid overlapping between vowels distributions, especially for mid/central vowels. This clearly confirms the validity of our proposed \texttt{VF RMSE} metric and reveals the limitations of sota ZS-TTS in modelling accents.

\vspace{-1.25em}
\section{Conclusions}
\vspace{-0.5em}

% In this paper, we propose to add transcription information, auxiliary highlight tasks, and meticulous screening for better subjective evaluation of accents; propose and verify that the pronunciation-related metrics (distances between vowel formants and PPGs) are reliable to objective evaluation of accents; and recommend using pronunciation, accent similarity, speaker similarity, and audio quality metrics and not using WER/CER/UTMOS for evaluating underrepresented accents. Our future work will aim at experimenting with different test designs (e.g.\ XMOS, XCMOS), more L1/L2 accents, and more synthesis systems, while exploring the  limitations of these metrics to ensure fair and inclusive evaluation of accent generation.

In this work, we enhance subjective accent evaluation by adding transcription information, auxiliary highlight tasks, and meticulous screening. Additionally, we demonstrate that pronunciation-related metrics like vowel formant distances and phonetic posteriorgrams distances, serve as reliable indicators for objective accent evaluation. Based on our findings, we recommend assessing underrepresented accents using pronunciation, accent similarity, speaker similarity, and audio quality metrics while avoiding WER, CER, and UTMOS, which may not accurately capture accent characteristics. Future work will explore alternative test designs (e.g.\ XMOS, XCMOS), a broader range of L1/L2 accents, and additional synthesis systems. We also aim to examine the limitations of these metrics, to ensure a fair and inclusive framework for accent evaluation.

% \begin{enumerate}
%     \item Findings \red{TBD}
%     \begin{enumerate}
%         \item DO consider doing these in subjective evaluation
%         \item DO NOT use these objective metrics
%         \item DO use these objective metrics
%     \end{enumerate}
%     \item Limitations
%     \begin{enumerate}
%         \item other test designs such as XMOS in subjective evaluation
%         \item accent-speaker entanglement in objective evaluation metric (e.g. speaker similarity, MCD)
%         \item extension to L2 accents (should we recruit L1 or L2 listeners)
%         \item proposed metrics extension to more models
%     \end{enumerate}
% \end{enumerate}

% \newpage
\vspace{-1.25em}
\section{Acknowledgements}
\vspace{-0.5em}
\ifinterspeechfinal
     This work was supported in part by the UKRI AI CDT in Responsible and Trustworthy in-the-world NLP (Grant EP/Y030656/1), UKRI CDT in NLP (Grant EP/S022481/1), School of Informatics, and School of Philosophy, Psychology \& Language Sciences, the University of Edinburgh.
\else
     [REDACTED FOR ANONYMITY]\\ \\ \\ \\
\fi

\bibliographystyle{IEEEtran}
\bibliography{mybib}

\begin{thebibliography}{10}
\providecommand{\url}[1]{#1}
\csname url@samestyle\endcsname
\providecommand{\newblock}{\relax}
\providecommand{\bibinfo}[2]{#2}
\providecommand{\BIBentrySTDinterwordspacing}{\spaceskip=0pt\relax}
\providecommand{\BIBentryALTinterwordstretchfactor}{4}
\providecommand{\BIBentryALTinterwordspacing}{\spaceskip=\fontdimen2\font plus
\BIBentryALTinterwordstretchfactor\fontdimen3\font minus \fontdimen4\font\relax}
\providecommand{\BIBforeignlanguage}[2]{{%
\expandafter\ifx\csname l@#1\endcsname\relax
\typeout{** WARNING: IEEEtran.bst: No hyphenation pattern has been}%
\typeout{** loaded for the language `#1'. Using the pattern for}%
\typeout{** the default language instead.}%
\else
\language=\csname l@#1\endcsname
\fi
#2}}
\providecommand{\BIBdecl}{\relax}
\BIBdecl

\bibitem{casanova2024xtts}
E.~Casanova, K.~Davis, E.~Gölge, G.~Göknar, I.~Gulea, L.~Hart, A.~Aljafari, J.~Meyer, R.~Morais, S.~Olayemi, and J.~Weber, ``{XTTS: a Massively Multilingual Zero-Shot Text-to-Speech Model},'' in \emph{Proc. Interspeech}, 2024, pp. 4978--4982.

\bibitem{chen2025neural}
S.~Chen, C.~Wang, Y.~Wu, Z.~Zhang, L.~Zhou, S.~Liu, Z.~Chen, Y.~Liu, H.~Wang, J.~Li, L.~He, S.~Zhao, and F.~Wei, ``{Neural Codec Language Models are Zero-Shot Text to Speech Synthesizers},'' \emph{IEEE/ACM TASLP}, vol.~33, pp. 705--718, 2025.

\bibitem{wang2025maskgct}
Y.~Wang, H.~Zhan, L.~Liu, R.~Zeng, H.~Guo, J.~Zheng, Q.~Zhang, X.~Zhang, S.~Zhang, and Z.~Wu, ``{Mask{GCT}: Zero-Shot Text-to-Speech with Masked Generative Codec Transformer},'' in \emph{ICLR}, 2025.

\bibitem{zhang2023towards}
M.~Zhang, X.~Zhou, Z.~Wu, and H.~Li, ``{Towards Zero-Shot Multi-Speaker Multi-Accent Text-to-Speech Synthesis},'' \emph{IEEE Signal Processing Letters}, vol.~30, pp. 947--951, 2023.

\bibitem{zhong2025accentbox}
J.~Zhong, K.~Richmond, Z.~Su, and S.~Sun, ``{AccentBox: Towards High-Fidelity Zero-Shot Accent Generation},'' in \emph{IEEE ICASSP}, 2025, pp. 1--5.

\bibitem{zhou2024multi}
X.~Zhou, M.~Zhang, Y.~Zhou, Z.~Wu, and H.~Li, ``{Multi-Scale Accent Modeling with Disentangling for Multi-Speaker Multi-Accent TTS Synthesis},'' \emph{arXiv preprint arXiv:2406.10844}, 2024.

\bibitem{zhou2024accented}
X.~Zhou, M.~Zhang, Y.~Zhou, Z.~Wu, and H.~Li, ``{Accented Text-to-Speech Synthesis With Limited Data},'' \emph{IEEE/ACM TASLP}, vol.~32, pp. 1699--1711, 2024.

\bibitem{liu2024controllable}
R.~Liu, B.~Sisman, G.~Gao, and H.~Li, ``{Controllable Accented Text-to-Speech Synthesis With Fine and Coarse-Grained Intensity Rendering},'' \emph{IEEE/ACM TASLP}, vol.~32, pp. 2188--2201, 2024.

\bibitem{melechovsky2024dart}
J.~Melechovsky, A.~Mehrish, B.~Sisman, and D.~Herremans, ``{DART}: Disentanglement of accent and speaker representation in multispeaker text-to-speech,'' in \emph{NeurIPS Workshops}, 2024.

\bibitem{badlani2023rad}
R.~Badlani, R.~Valle, K.~J. Shih, J.~F. Santos, S.~Gururani, and B.~Catanzaro, ``{RAD-MMM: Multilingual Multiaccented Multispeaker Text To Speech},'' in \emph{Proc. Interspeech}, 2023, pp. 626--630.

\bibitem{zhang2022accentspeech}
Y.~Zhang, Z.~Wang, P.~Yang, H.~Sun, Z.~Wang, and L.~Xie, ``{AccentSpeech: Learning Accent from Crowd-sourced Data for Target Speaker TTS with Accents},'' in \emph{IEEE ISCSLP}, 2022, pp. 76--80.

\bibitem{deja2023diffusion}
K.~Deja, G.~Tinchev, M.~Czarnowska, M.~Cotescu, and J.~Droppo, ``{Diffusion-based Accent Modelling in Speech Synthesis},'' in \emph{Proc. Interspeech}, 2023, pp. 5516--5520.

\bibitem{yang2023parameter}
L.-J. Yang, C.-H.~H. Yang, and J.-T. Chien, ``{Parameter-Efficient Learning for Text-to-Speech Accent Adaptation},'' in \emph{Proc. Interspeech}, 2023, pp. 4354--4358.

\bibitem{ding2022accentron}
S.~Ding, G.~Zhao, and R.~Gutierrez-Osuna, ``{Accentron: Foreign Accent Conversion to Arbitrary Non-native Speakers Using Zero-shot Learning},'' \emph{Computer Speech \& Language}, vol.~72, p. 101302, 2022.

\bibitem{quamer2022zero}
W.~Quamer, A.~Das, J.~Levis, E.~Chukharev-Hudilainen, and R.~Gutierrez-Osuna, ``{Zero-Shot Foreign Accent Conversion without a Native Reference},'' in \emph{Proc. Interspeech}, 2022, pp. 4920--4924.

\bibitem{zhou2023tts}
Y.~Zhou, Z.~Wu, M.~Zhang, X.~Tian, and H.~Li, ``{TTS-Guided Training for Accent Conversion Without Parallel Data},'' \emph{IEEE Signal Processing Letters}, vol.~30, pp. 533--537, 2023.

\bibitem{bai2024diffusion}
Q.~Bai, S.~Wang, Z.~Liu, M.~Zhang, W.~Rao, Y.~Wang, and H.~Li, ``{Diffusion-Based Method with TTS Guidance for Foreign Accent Conversion},'' in \emph{IEEE ISCSLP}, 2024, pp. 284--288.

\bibitem{chen2024transfer}
X.~Chen, J.~Pei, L.~Xue, and M.~Zhang, ``{Transfer the Linguistic Representations from TTS to Accent Conversion with Non-Parallel Data},'' in \emph{IEEE ICASSP}, 2024, pp. 12\,501--12\,505.

\bibitem{jia2023zero}
D.~Jia, Q.~Tian, K.~Peng, J.~Li, Y.~Chen, M.~Ma, Y.~Wang, and Y.~Wang, ``{Zero-Shot Accent Conversion using Pseudo Siamese Disentanglement Network},'' in \emph{Proc. Interspeech}, 2023, pp. 5476--5480.

\bibitem{jin2023voice}
M.~Jin, P.~Serai, J.~Wu, A.~Tjandra, V.~Manohar, and Q.~He, ``{Voice-Preserving Zero-Shot Multiple Accent Conversion},'' in \emph{IEEE ICASSP}, 2023, pp. 1--5.

\bibitem{jia2024convert}
Z.~Jia, H.~Xue, X.~Peng, and Y.~Lu, ``{Convert and Speak: Zero-shot Accent Conversion with Minimum Supervision},'' in \emph{Proc. ACM MM}, 2024, p. 4446–4454.

\bibitem{inoue2025macst}
S.~Inoue, S.~Wang, W.~Wang, P.~Zhu, M.~Bi, and H.~Li, ``{MacST: Multi-Accent Speech Synthesis via Text Transliteration for Accent Conversion},'' in \emph{IEEE ICASSP}, 2025, pp. 1--5.

\bibitem{saeki2022utmos}
T.~Saeki, D.~Xin, W.~Nakata, T.~Koriyama, S.~Takamichi, and H.~Saruwatari, ``{UTMOS: UTokyo-SaruLab System for VoiceMOS Challenge 2022},'' in \emph{Proc. Interspeech}, 2022, pp. 4521--4525.

\bibitem{mittag2020deep}
G.~Mittag and S.~Möller, ``{Deep Learning Based Assessment of Synthetic Speech Naturalness},'' in \emph{Proc. Interspeech}, 2020, pp. 1748--1752.

\bibitem{ren2021fastspeech}
Y.~Ren, C.~Hu, X.~Tan, T.~Qin, S.~Zhao, Z.~Zhao, and T.-Y. Liu, ``Fastspeech 2: Fast and high-quality end-to-end text to speech,'' in \emph{ICLR}, 2021.

\bibitem{munro1995foreign}
M.~J. Munro and T.~M. Derwing, ``{Foreign Accent, Comprehensibility, and Intelligibility in the Speech of Second Language :earners},'' \emph{Language Learning}, vol.~45, no.~1, pp. 73--97, 1995.

\bibitem{markl2023everyone}
N.~Markl and C.~Lai, ``{Everyone has an Accent},'' in \emph{Proc. Interspeech}, 2023, pp. 4424--4427.

\bibitem{camp2023mos}
J.~Camp, T.~Kenter, L.~Finkelstein, and R.~Clark, ``{MOS vs. AB: Evaluating Text-to-Speech Systems Reliably Using Clustered Standard Errors},'' in \emph{Proc. Interspeech}, 2023, pp. 1090--1094.

\bibitem{wells2024experimental}
D.~Wells, A.~L. {Aldana Blanco}, C.~Valentini, E.~Cooper, A.~Pine, J.~Yamagishi, and K.~Richmond, ``{Experimental Evaluation of MOS, AB and BWS Listening Test Designs},'' in \emph{Proc. Interspeech}, 2024, pp. 2695--2699.

\bibitem{liu2024a}
S.~Liu, M.~Babel, and J.~Zhu, ``{A Comparison of Voice Similarity through Acoustics, Human Perception and Deep Neural Network (DNN) Speaker Verification Systems},'' in \emph{Proc. Interspeech}, 2024, pp. 3674--3678.

\bibitem{sanabria2023the}
R.~Sanabria, N.~Bogoychev, N.~Markl, A.~Carmantini, O.~Klejch, and P.~Bell, ``{The Edinburgh International Accents of English Corpus: Towards the Democratization of English ASR},'' in \emph{IEEE ICASSP}, 2023, pp. 1--5.

\bibitem{le2024limits}
S.~Le~Maguer, S.~King, and N.~Harte, ``{The limits of the Mean Opinion Score for speech synthesis evaluation},'' \emph{Computer Speech \& Language}, vol.~84, p. 101577, 2024.

\bibitem{huckvale2004accdist}
M.~Huckvale, ``{ACCDIST: a Metric for Comparing Speakers' Accents},'' in \emph{Proc. Interspeech}, 2004, pp. 29--32.

\bibitem{churchwell2024high}
C.~Churchwell, M.~Morrison, and B.~Pardo, ``{High-Fidelity Neural Phonetic Posteriorgrams},'' in \emph{IEEE ICASSP Workshops}, 2024, pp. 823--827.

\bibitem{gutierrez2021location}
E.~Gutierrez, P.~Oplustil-Gallegos, and C.~Lai, ``{Location, Location: Enhancing the Evaluation of Text-to-Speech synthesis using the Rapid Prosody Transcription Paradigm},'' in \emph{ISCA SSW}, 2021, pp. 25--30.

\bibitem{zuluagagomez2023commonaccent}
J.~Zuluaga-Gomez, S.~Ahmed, D.~Visockas, and C.~Subakan, ``{CommonAccent: Exploring Large Acoustic Pretrained Models for Accent Classification Based on Common Voice},'' in \emph{Proc. Interspeech}, 2023, pp. 5291--5295.

\bibitem{chen2022WavLMLargeScaleSelfSupervised}
S.~Chen, C.~Wang, Z.~Chen, Y.~Wu, S.~Liu, Z.~Chen, J.~Li, N.~Kanda, T.~Yoshioka, X.~Xiao, J.~Wu, L.~Zhou, S.~Ren, Y.~Qian, Y.~Qian, J.~Wu, M.~Zeng, X.~Yu, and F.~Wei, ``{WavLM}: {Large}-{Scale} {Self}-{Supervised} {Pre}-{Training} for {Full} {Stack} {Speech} {Processing},'' \emph{IEEE Journal of Selected Topics in Signal Processing}, vol.~16, no.~6, pp. 1505--1518, Oct. 2022.

\bibitem{zhang2024amphion}
X.~Zhang, L.~Xue, Y.~Gu, Y.~Wang, J.~Li, H.~He, C.~Wang, S.~Liu, X.~Chen, J.~Zhang, Z.~Fang, H.~Chen, T.~Y. Tang, L.~Zou, M.~Wang, J.~Han, K.~Chen, H.~Li, and Z.~Wu, ``{Amphion: an Open-Source Audio, Music, and Speech Generation Toolkit},'' in \emph{IEEE SLT}, 2024, pp. 879--884.

\bibitem{wells1982accents}
J.~C. Wells, \emph{{Accents of English: Volume 1}}.\hskip 1em plus 0.5em minus 0.4em\relax Cambridge University Press, 1982, vol.~1.

\bibitem{mcauliffe2017MontrealForcedAligner}
M.~McAuliffe, M.~Socolof, S.~Mihuc, M.~Wagner, and M.~Sonderegger, ``{Montreal Forced Aligner: Trainable Text-Speech Alignment Using Kaldi},'' in \emph{Proc. Interspeech}, 2017, pp. 498--502.

\bibitem{barreda2021fast}
S.~Barreda, ``{Fast Track: Fast (nearly) Automatic Formant-tracking Using Praat},'' \emph{Linguistics Vanguard}, vol.~7, no.~1, p. 20200051, 2021.

\bibitem{kong2020hifigan}
J.~Kong, J.~Kim, and J.~Bae, ``{HiFi-GAN: Generative Adversarial Networks for Efficient and High Fidelity Speech Synthesis},'' in \emph{NeurIPS}, H.~Larochelle, M.~Ranzato, R.~Hadsell, M.~Balcan, and H.~Lin, Eds., vol.~33, 2020, pp. 17\,022--17\,033.

\end{thebibliography}

\end{document}